\documentclass[twocolumn,amssymb,fleqn]{revtex4} %fleqn: make all eqns left alighed  ,showpacs
\usepackage{epsfig,amssymb,amsmath,graphicx,hyperref  }  % pdfpages and graphicx are added   hyperref(set hyperlink for content)

\begin{document}

\title{Curvature-driven stability of defects in nematic textures over spherical
disks}
\author{Xiuqing Duan and Zhenwei Yao}
\email{zyao@sjtu.edu.cn}
\affiliation{School of Physics and Astronomy, and Institute of Natural
Sciences, Shanghai Jiao Tong University, Shanghai 200240, China}
\begin{abstract}
Stabilizing defects in liquid-crystal systems is crucial for many physical
processes and applications ranging from functionalizing liquid-crystal textures
to recently reported command of chaotic behaviors of active matters. In this
work, we perform analytical calculations to study the curvature driven
stability mechanism of defects based on the isotropic nematic disk model that
is free of any topological constraint. We show that in a growing spherical disk
covering a sphere the accumulation of curvature effect can prevent typical +1
and +1/2 defects from forming boojum textures where the defects are repelled to
the boundary of the disk. Our calculations reveal that the movement of the
equilibrium position of the +1 defect from the boundary to the center of the
spherical disk occurs in a very narrow window of the disk area, exhibiting the
first-order phase-transition-like behavior. For the pair of +1/2 defects by
splitting a +1 defect, we find the curvature driven alternating repulsive and
attractive interactions between the two defects. With the growth of the
spherical disk these two defects tend to approach and finally recombine towards
a +1 defect texture. The sensitive response of defects to curvature and the
curvature driven stability mechanism demonstrated in this work in nematic disk
systems may have implications towards versatile control and engineering of
liquid crystal textures in various applications.
\end{abstract}
%\pacs{61.72.-y, 61.46.Bc, 61.72.J-, 61.72.Bb}
\maketitle

\section{Introduction}

Functionalizing the rich variety of self-assembled liquid-crystal (LC)
structures represents a trend in LC research~\cite{bisoyi2011liquid,
alexander2012colloquium, umadevi2013large,urbanski2017liquid}. Confining LCs in
various geometries in the form of 
droplets~\cite{lopez2011frustrated,pairam2013stable,yamamoto2016chirality},
shells~\cite{fernandez2007novel, lopez2009topological,liang2012towards} and
fibers~\cite{yoshino2010three, fleischmann2014liquid} using modern
microfluidic technology and characterization methods opens the prospect of many
application opportunities, and brings new scientific problems related to the
creation and engineering of complex director arrangements~\cite{de1995physics,
lopez2011drops, sengupta2012functionalization, urbanski2017liquid}. LC
textures can be strongly affected by the distribution and type of topological
defects, which are singularities in the otherwise continuous LC director
field~\cite{kleman1983points, chaikin2000principles, nelson2002defects}.  The
extraordinary responsiveness of LC makes the manipulation of defects a
challenge in applications. Stabilizing defects in two-dimensional LC systems is
directly related to arrangement of LC
textures~\cite{kleman1983points,de1995physics,mbanga2014simulating,darmon2016topological},
fabrication of controllable valency in colloid-LC-based artificial
atoms~\cite{nelson2002toward, gharbi2013microparticles, koning2016spherical},
modulation of coupled geometries where LC lives~\cite{frank2008defects,
xing2012morphology,chen2012threading,ramakrishnan2013membrane,pismen2014metric,
mostajeran2015curvature, leoni2017defect}, and relevant applications in active
matter systems~\cite{sanchez2012spontaneous,zhou2014living,keber2014topology,
lavrentovich2016active, peng2016command}.  A prototype model to study the
stability mechanism of defects in LC is the isotropic two-dimensional LC disk
model with a single elastic
constant~\cite{langer1986textures,rudnick1995shape,pettey1999stability}.  In a
flat freestanding LC disk, defects tend to move swiftly to the boundary to
form a boojum texture, which is a two-dimensional version of its namesake in
superfluid helium-3~\cite{bhattacharyya1977stability, stein1978boojums,
langer1986textures}. A ``virtual boojum" texture with a topological defect
outside the sample has been predicted in planar circular LC domains by Langer
and Sethna ~\cite{langer1986textures}, and it has been found to be a local
energy extremal~\cite{rudnick1995shape, pettey1999stability}.  Sufficiently
strong pinning boundary conditions can stabilize a defect within a circular LC
domain~\cite{langer1986textures,rudnick1995shape,pettey1999stability}.

Exploring other stability mechanisms of defects in LC samples in addition to
imposing boundary conditions constitutes an underlying scientific problem
towards versatile control and engineering of LC director arrangement. Confining
LC over spherical surfaces can generate various regularly arranged stable
defect patterns~\cite{vitelli2006nematic,
fernandez2007novel,shin2008topological,bates2008nematic, bowick2009two,
xing2012morphology,sevc2012defect,koning2016spherical}.   Vitelli and Nelson
have studied two-dimensional nematic order coating frozen surfaces of spatially
varying Gaussian curvature, and found
the instability of a smooth ground-state texture to the generation of a single
defect using free boundary conditions~\cite{vitelli2004defect}. These results
of LC order on closed spheres and topographies with varying curvature show that
curvature suffices to provide a stability mechanism for defects even
without imposing any pinning boundary condition. However, for LC order on a
closed sphere, it is unknown to what extent the appearance of defects is
energetically driven, while they must appear as a consequence of the spherical
topology. To remove the topological constraint, we study nematic order, the
simplest LC order, on a spherical disk. Here we emphasize that, due to the
fundamentally distinct topologies of sphere and disk, the appearance
of defects on spherical disks is not topologically required; the emergence of defects therein is purely
geometrically driven. According to the continuum elasticity theory of
topological defects in either LC or crystalline order, the stress caused by
defects can be partially screened by Gaussian
curvature~\cite{chaikin2000principles, nelson2002defects, bowick2009two,
koning2016crystals}. Therefore, one expects the appearance of defects on a
sufficiently curved spherical disk. It is of
interest to identify the transition point for a defect to depart from the
boundary of the disk, and illustrate the nature of the
transition by clarifying questions such as: Will the defect move rapidly or
gradually with the accumulation of curvature effect? Will the defect split as
the nematic texture becomes more and more frustrated by the curvature? Once
split, will the resulting defects become stable on the spherical disk?

We perform analytical calculations based on the isotropic nematic disk model to
address these fundamental problems. This theoretical model may be realized
experimentally in Langmuir
monolayers~\cite{knobler1992phase,fang1997boojums,pettey1999stability,gupta2012liquid}
and liquid-crystal films~\cite{langer1986textures,xue1992phase}  deposited at
the surface of water droplets whose curvature is controllable by tuning the
droplet size~\cite{urbanski2017liquid}.  Flat space experiments in these
two-dimensional monolayer systems at air-water interface have revealed stable
liquid-crystal phases~\cite{knobler1992phase, xue1992phase,gupta2012liquid}.

In this work, we first discuss the two instability modes
of a +1 defect over a flat disk, either sliding to the boundary or splitting
to a pair of +1/2 defects. By depositing the nematic order over a spherical
surface, we analytically show that bending deformation of a director field is
inevitable {\it everywhere}, which implies the appearance of defects to release
the curvature-driven stress.  By comparing a flat and a spherical nematic disk
of the same area, both containing a +1 defect at the center, we derive for the
analytical expression for the difference of the Frank free energy, and show
that the spherical disk always has higher energy.  However, when the +1 defect
deviates from the center of the disk, the free energy curves become
qualitatively different for flat and spherical disks when the disk area exceeds
some critical value. Specifically, the equilibrium position of the +1 defect
rapidly moves from the boundary to the center of the spherical disk in a narrow
window of the disk area, exhibiting the first-order phase-transition-like
behavior. For the pair of +1/2 defects by splitting a +1 defect, we further
show the curvature-driven alternating repulsive and attractive interactions
between the two defects. When the spherical cap occupies more area over the
sphere, the pair of +1/2 defects tend to approach until merging to a +1
defect texture. The recombination of the pair of +1/2 defects into a +1 defect
is consistent with the result of the +1 defect case. These results demonstrate
the fundamentally distinct scenario of defects in a spherical disk from that on
a planar disk. We also briefly discuss the cases of nematic order on hyperbolic
disks. In this work, the demonstrated distinct energy landscape of LC defects
created by curvature is responsible for the stability of defects, and may have
implications in the design of LC textures with the dimension of curvature.

\section{Model and Method}

In the continuum limit, the orientations of liquid-crystal molecules lying over
a disk are characterized by a director field $\mathbf{n}(\mathbf{x})$ that is
defined at the associated tangent plane at $\mathbf{x}$.  The equilibrium
nematic texture is governed by minimizing the Frank free
energy~\cite{chaikin2000principles}
\begin{equation} F= \int_{D} f dA + \lambda (\mathbf{n}^2-1),
\end{equation}
where the integration is over the disk $D$. The Frank
free energy density
\begin{equation} f=\frac{1}{2}K_1(\textrm{div} \,
  \mathbf{n})^2+\frac{1}{2}K_3(\mathbf{n} \times \textrm{curl} \, \mathbf{n})^2,\label{Frank}
  \end{equation}
where $K_1$ and $K_3$ are the splay and bending rigidities, respectively.  The
Lagrange multiplier $\lambda$ is introduced to implement the constraint of
$\mathbf{n}\cdot \mathbf{n}=1$. In general, $\lambda$ is a function of
coordinates.  The twist term $(\mathbf{n} \cdot \textrm{curl} \, \mathbf{n})^2$
vanishes in nematics confined on a sphere (see Appendix B). Equation (\ref{Frank}) has been widely used to
analyze the deformation in nematic phases.  For nematics on curved surfaces,
the operators of divergence and curl in the Frank free energy are promoted to
be defined on the curved manifold and carry the information of curvature. Note
that the curl operator relies on the extrinsic geometry of the surface
~\cite{napoli2012extrinsic}. Note that the Frank free energy model in
Eq. (\ref{Frank}) describes the distortion free energy of uniaxial nematics. A
formalism based on the tensorial nematic order parameter has been proposed to
characterize the distortion of both uniaxial and biaxial
nematics and defects therein~\cite{kralj2011curvature, rosso2012parallel}.

We work in the approximation of isotropic elasticity with $K_1 = K_3$. Under
such an approximation, one can show that the free energy is invariant under the
local rotation of the director field by any angle, whether the disk is planar or
curved (see Appendix A). In other words, the energy degeneracy of the system
becomes infinite when $K_1 = K_3$. Such configurational symmetry is broken when
the ratio $K_1/K_3$ is deviated from unity. While the states selected by
the differential in the values for $K_1$ and $K_3$ are of interest in other
contexts such as in the ground states of spherical nematics~\cite{bowick2009two},
here we work in the isotropic regime to highlight the curvature effect
of substrates on the configuration of nematics.

The general Euler-Lagrange equation of the Frank free energy on
a curved surface $\mathbf{x}(u^1, u^2)$ is
\begin{eqnarray}
  \partial_j
\frac{\partial f}{\partial (\frac{\partial n_i}{\partial
u^j})}+\frac{\partial_j \sqrt g}{\sqrt g} \frac{\partial f}{\partial
(\frac{\partial n_i}{\partial u^j})} - \frac{\partial f}{\partial n_i} =
-\lambda n_i,\label{EL_general}
\end{eqnarray}
where $i, j=1,2$, and $g$ is the determinant of the metric tensor. The second term in
Eq.(\ref{EL_general}) is due to the spatially varying $g$. The nematic textures studied in this work are 
solutions to Eq.(\ref{EL_general}).

To characterize defects that are named disclinations in a two-dimensional
director field, we perform integration of the orientation $\theta$ of the
director $\mathbf{n}$ with respect to any local reference frame along any
closed loop $\Gamma$:
\begin{eqnarray}
  \oint_{\Gamma} d\theta = k\pi,
\end{eqnarray}
where $k$ is nonzero if $\Gamma$ contains a defect.  Unlike in a vector field
where $k$ can only be integers, two-dimensional nematics supports both
integer and half-integer disclinations due to the apolarity of liquid-crystal
molecules, i.e., $\mathbf{n}\equiv -\mathbf{n}$.

\section{Results and discussion}

\begin{figure}
\includegraphics[width=8.5cm]{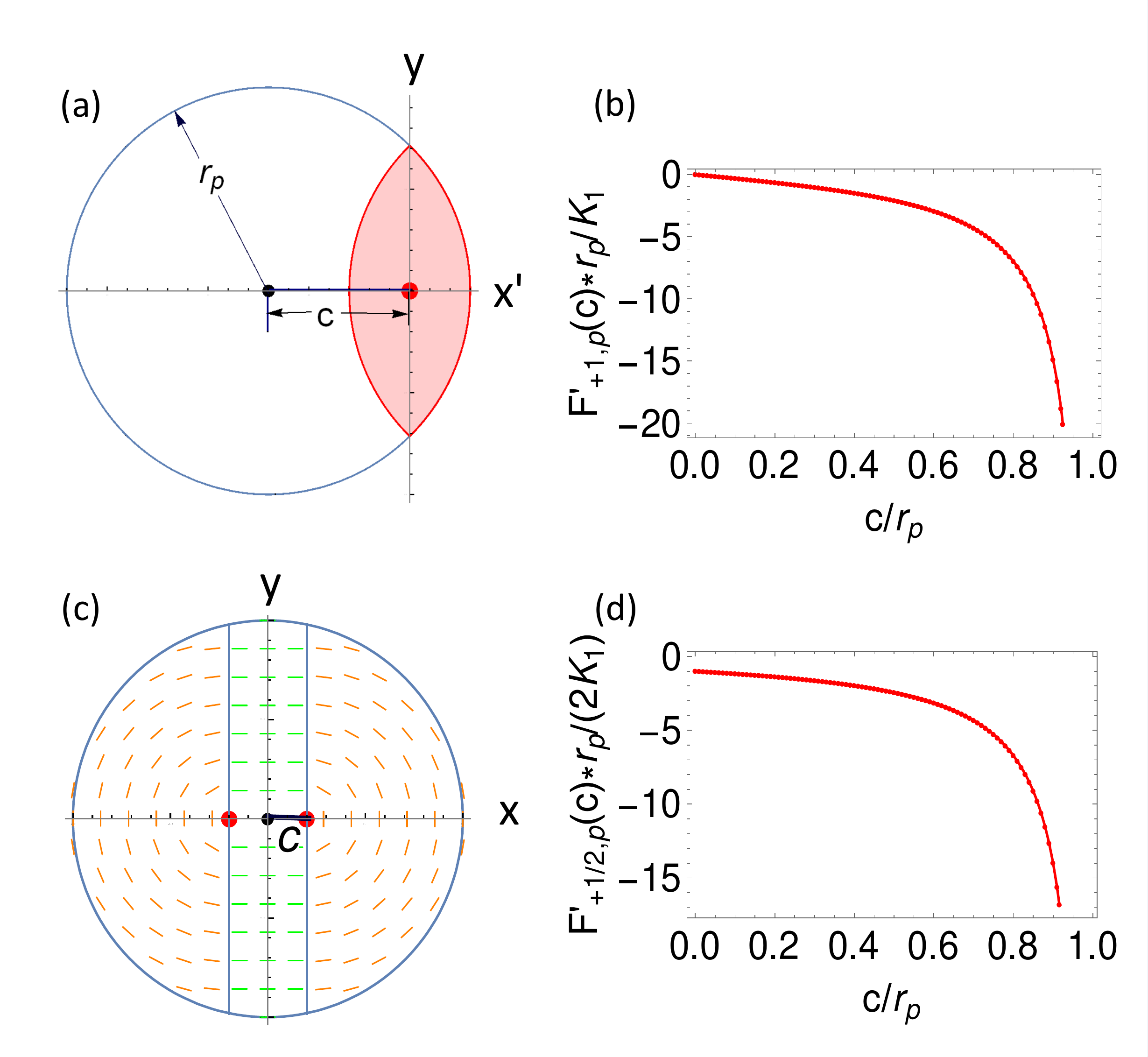}

\caption{ The configurations and energetics of a single +1 defect (a,b) and a
  pair of +1/2 defects (c,d) on a planar disk. The defects are represented by red
    dots. The pair of +1/2 defects in (c) are constructed out of a +1 defect by
    inserting a uniform direct field between (indicated by green lines). The negative derivative of the Frank free energy $F(c)$ with respect to $c$
    (the position of the defects) indicates that defects tend to slide to the
    boundary of a planar disk.
  }
  \label{int_domain}
\end{figure}

We first discuss the case of nematics on a planar disk. 
It is straightforward to identify the following solution to the Euler-Lagrange equation:
\begin{equation} \mathbf{n}=
  \cos(\varphi+\theta_0) \mathbf{e}_1+  \sin(\varphi+\theta_0)\mathbf{e}_2, \label{aster}
\end{equation}
where $\varphi = \textrm{arctan}(y/x)$ is the polar angle, $\theta_0$ is a
constant, and $\mathbf{e}_i$ is the unit basis vector in Cartesian coordinates.
The strength of the defect located at the origin of the coordinates is +1. The associated Lagrange multiplier is $\lambda=K_1/(x^2+y^2)$.
The contributions to the splay and bending terms in the
free energy density are $K_1 \textrm{cos}^2 \theta_0/[2(x^2+y^2)]$ and
$K_3 \textrm{sin}^2 \theta_0/[2(x^2+y^2)]$, respectively. When
$\theta_0$ increases from 0 to $\pi/2$, the +1 defect transforms from
the radial (pure splay) to the azimuthal (pure bending) configurations. In this
process, the sum of the splay and bending energies is an invariant under the
isotropic elasticity approximation. The total free energy of the configuration
in Eq.(\ref{aster}) is
\begin{eqnarray}
  F_{+1, p} = \frac{K_1}{2} \iint_{x^2+y^2 \leq r_p^2} \frac{1}{x^2+y^2}
  dxdy,\label{Fp}
\end{eqnarray}
where $r_p$ is the radius of the planar disk.

We show that the +1 defect at the center of the planar disk in Eq.(\ref{aster})
is unstable and tends to slide to the boundary of the disk. For simplicity, we
employ free boundary condition. Consider a +1 defect like in Eq.(\ref{aster})
at $(c, 0)$, where $c \leq r_p$.  Its free energy is
\begin{eqnarray}\label{}
  F_{+1,p}(c)=\frac{K_1}{2} \iint_{x^2+y^2 \leq r_p^2} \frac{1}{(x-c)^2+y^2} dxdy.
\end{eqnarray}
To avoid the singularity point at $(c, 0)$ in the evaluation for $F_{+1,p}(c)$, we
take the derivative of $F_{+1,p}(c)$ with respect to $c$. Physically, this procedure returns
the force on the defect. While the free energy may diverge, a physical force
must be finite. After some calculation, we have
\begin{eqnarray}\label{}
  F'_{+1,p}(c)=K_1 \iint_{x^2+y^2 \leq r_p^2} \frac{x-c}{[(x-c)^2+y^2]^2} dxdy \nonumber \\
  = K_1 \iint_{(x'+c)^2+y^2 \leq r_p^2} \frac{x'}{(x'^2+y^2)^2} dx'dy,
\end{eqnarray}
where variable substitution is applied in the last equality.  The integral domain is shown in Fig.~\ref{int_domain}(a). The defect is located at $x'=0$
(i.e., $x = c$) and $y=0$. We see that the integration in the red region returns zero, since
the integrand $x'/(x'^2+y^2)^2$ is an odd function of $x'$. In the rest
region where $x'<0$, the integrand is negative. Therefore,
$F'_{+1,p}(c)$ is negative when the defect is deviated from the center of the disk. $F'_{+1,p}(c=0)=0$. In other words, once deviated from the center of the disk, the defect will slide to the
boundary to reduce the free energy of the system. Figure~\ref{int_domain}(b) shows the numerical result on the dependence of
$F'_{+1,p}(c)$ on $c/r_p$.

An alternative instability mode of the central +1 defect in the planar disk is
to split into two +1/2 defects. Such a process may occur when the
interaction energy of the two repulsive +1/2 defects dominates over the core energy of the defects. To analyze the
energetics of the +1/2 defects, we construct the director field containing two
+1/2 defects by cutting and moving apart an azimuthal configuration as shown in
Fig.1(c), where the +1/2 defects are represented by red dots. The region
between the two half azimuthal configurations is filled with a uniform director
field. The Frank free energy of such a configuration is
\begin{eqnarray}
  F_{+1/2,p}(c)=2K_1
\int_0^{r_p-c}dx\int_0^{\sqrt{r_p^2-(x+c)^2}}dy \frac{1}{x^2+y^2},
  \label{two_half}
  \end{eqnarray}
where the separation between the two defects is $2c$. $F'(0)=-2K_1/r_p<0$. In Fig.~\ref{int_domain}(d), we plot $F'_{+1/2,p}(c)$
versus $c$. The negative
sign indicates the repulsive nature of the two +1/2 defects. The resulting +1/2
defects are ultimately pushed to the boundary of the disk under the repulsive
interaction.

\begin{figure}
\includegraphics[width=8.5cm, bb=72 253 764 496]{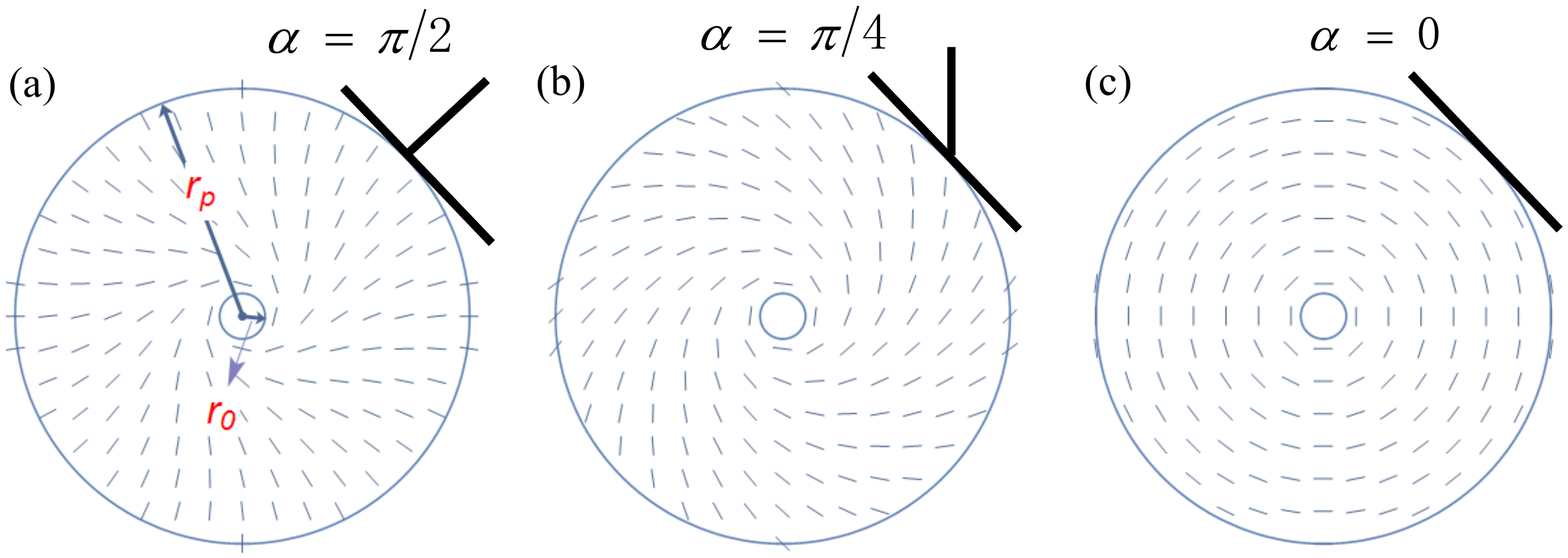}
\caption{Spiral +1 defect patterns subject to typical pinning boundary conditions in a planar nematic disk. 
}
\label{spiral}
\end{figure}

In the preceding discussions, we employ the free boundary condition where
directors at the boundary do not have preferred orientations. Another important
class of boundary condition is to fix the orientation of the molecules at the
boundary. Homeotropic and planar liquid-crystal samples are two typical cases,
where the directors are perpendicular and parallel to the boundary,
respectively. Imposing these pinning boundary conditions over the aster
configuration can lead to spiral deformations~\cite{de1995physics}. Note that a
recent study has demonstrated a dynamic consequence of the radial-to-spiral
transition of a +1 defect pattern in the system of swimming bacteria in a
liquid-crystal environment~\cite{peng2016command}.  It is observed that the swimming
mode of bacteria changes from bipolar to unipolar when the +1 defect pattern
becomes spiral. For the general pinning boundary condition that the angle between $\mathbf{n}$ and
the tangent vector at the boundary is $\alpha$ ($\alpha \in [0, \pi/2]$), we obtain the solution to
Eq.(\ref{EL_general}): 
\begin{eqnarray}
  (n_r,n_{\varphi})=\left( \textrm{sin} \theta(\alpha),\textrm{cos} \theta(\alpha) \right),\label{spiral_solution}
\end{eqnarray}
where $n_r$ and $n_{\varphi}$ are the components of $\mathbf{n}$ in polar coordinates $(r, \varphi)$,
$\theta(\alpha) = \alpha\ \textrm{ln}(r/r_0)/\textrm{ln}(r_p/r_0)$, $r_p$ and
$r_0$ are the outer and inner radius of the planar disk as shown in Fig.~\ref{spiral}. The magic spiral
solution in Ref.~\cite{de1995physics} is a special case of $\alpha=\pi/2$. The associated
Lagrange multiplier is $\lambda=(K_1/r^2) \{1+\alpha^2/[\textrm{ln}
(r_p/r_0)]^2 \}$.
The configuration of the solution in Eq.(\ref{spiral_solution}) is plotted in Fig.~\ref{spiral}.
The originally straight radial lines deform to spiral
curves to satisfy the boundary condition. The Frank free energy of the spiral configuration is
\begin{equation}\label{}
  F_{\rm{spiral}}(\alpha)=\left( 1+\left(\frac{\alpha}{\textrm{ln} \frac{r_p}{r_0}}\right)^2
  \right) F_{+1, p},\label{F_spiral}
\end{equation}
where $F_{+1, p}$ is the free energy of an aster configuration in a planar disk
given in Eq.(\ref{Fp}).  Eq.(\ref{F_spiral}) shows that the boundary
effect does not enter the integral of $F_{+1, p}$. The energy cost
associated with the spiral deformation conforms to a quadratic law with
respect to the angle $\alpha$. And its dependence on the size of disk is
relatively weak in a logarithm relation.

Now we discuss two-dimensional nematic texture confined on spherical disks.
Consider a director field $\mathbf{n}$ on a sphere $\mathbf{n} =
n_1(\theta,\varphi)\mathbf{e}_{\theta} +
n_2(\theta,\varphi)\mathbf{e}_{\varphi}$, where 
$\mathbf{e}_{\theta}$ and $\mathbf{e}_{\varphi}$ are the unit tangent vectors
in spherical coordinates. $\theta$ and $\varphi$ are the polar and azimuthal
angles, respectively. 
We first show that on spherical geometry a director field without any splay and
bending deformations is impossible. Topology of the two-dimensional sphere
dictates that a harmonic vector field on a sphere is impossible~\cite{Nakahara}. A vector field is called harmonic if it is
divergence-free, irrotational, and tangent to the spherical surface. A
director field is a vector field with the extra constraints of $|\mathbf{n}|=1$
and $\mathbf{n} \equiv -\mathbf{n}$. Therefore, it is a topological requirement
that one cannot completely eliminate both bending and splay deformations in a
director field living on a sphere.

In addition to the above global analysis, we will further show that an
irrotational director field is impossible {\it at any point} on a sphere. In other words, bending
of a director field is inevitable {\it everywhere} on a spherical surface. We first present the
general expressions for the divergence and curl of a director field over a
smooth surface:  $\textrm{div}\ \mathbf{n} =
\frac{1}{\sqrt{g}} \partial_i(\sqrt{g} n_i)$ and  $\textrm{curl}\ 
\textbf{n} = (\star d \textbf{n}^{\flat})^{\sharp}$, where $\star$ is the Hodge dual,
$\scriptstyle\flat$ and $\scriptstyle\sharp$ are the musical isomorphisms, $d$ is exterior derivative (see Appendix B). Applying these
expressions on a sphere, we have
\begin{eqnarray}
  \textrm{div}\ \mathbf{n} = \frac{\textrm{cos} \theta}{\textrm{sin} \theta}
  \frac{n_1}{R} + \frac{1}{R} \frac{\partial n_1}{\partial \theta} + \frac{1}{R
  \textrm{sin}\theta} \frac{\partial n_2}{\partial \varphi},\label{div_sphere}
\end{eqnarray}
and
\begin{eqnarray}
\textrm{curl} \, \textbf{n} = \frac{1}{R \textrm{sin} \theta} (-\frac{\partial
  n_1}{\partial \varphi} + n_2 \textrm{cos} \theta + \textrm{sin} \theta
  \frac{\partial n_2}{\partial \theta}) \mathbf{e_{r}} \nonumber \\
 - \frac{n_2}{R} \mathbf{e_{\theta}} + \frac{n_1}{R} \mathbf{e_{\varphi}} ,  \label{curl_sphere}
\end{eqnarray}
where $\mathbf{e_{r}}$ is the unit normal vector. According to
Eq.(\ref{curl_sphere}), we clearly see that at least one of the last two terms
must be nonzero. In contrast, Eq.(\ref{div_sphere}) shows that a divergence
free director field with vanishing splay deformation without any bend
deformation is possible. The simplest example is the direction field with only the
azimuthal component: $\mathbf{n} = \mathbf{e}_{\varphi}$. Such a director field
is divergence free but with bending deformation. Note that in the calculation for the curl of the
director field, we use the condition that the sphere is embedded in
three-dimensional Euclidean space. The divergence of the director field does
not depend on how the sphere is embedded in the Euclidean space. One can check that
the twist term $(\mathbf{n} \cdot \textrm{curl} \, \mathbf{n})^2 = 0$.

The stability analysis of defects in nematic textures over spherical
disks is based on the following expression for the Frank free energy density in spherical
coordinates:
\begin{eqnarray}
  f&=&\frac{K_1}{2R^2}  (n_1 \frac{\textrm{cos} \theta}{\textrm{sin}
  \theta}+\frac{\partial n_1}{\partial \theta}+\frac{1}{\textrm{sin} \theta}
  \frac{\partial n_2}{\partial \varphi})^2  +\frac{K_3}{2R^2}
  \nonumber  \\  &&\times [1+\frac{1}{\textrm{sin}^2 \theta} (n_2 \textrm{cos} \theta + \textrm{sin} \theta
  \frac{\partial n_2}{\partial \theta}-\frac{\partial n_1}{\partial
  \varphi})^2 ], \label{Fsphere}
\end{eqnarray}
Note that the first term $K_3/(2R^2)$ in the bending part represents the
irremovable bending deformation of a director field over spherical substrates.
This term vanishes in the limit of $R\rightarrow \infty$.  One can check that
for a divergence-free director field $\mathbf{n} = \mathbf{e}_{\varphi}$,
  $f=K_1/(2R^2\sin^2\theta)$.  The singularities at $\theta=0$ and $\theta=\pi$
  correspond to the two +1 defects at the north and south poles.

\begin{figure}
\includegraphics[width=8cm, bb=89 242 511 568]{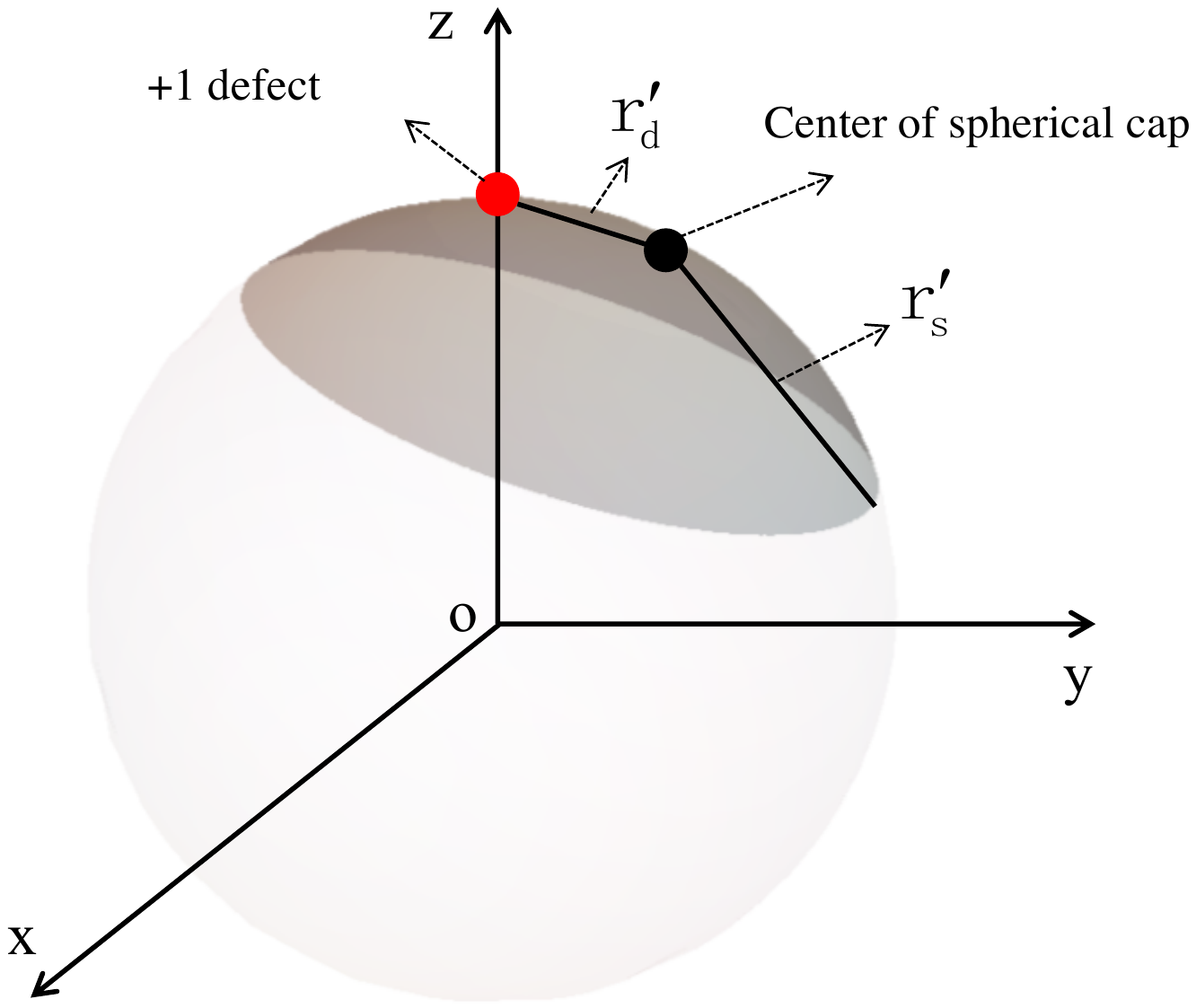}
\caption{Schematic plot of a +1 defect over a spherical cap. The relative
position the defect is characterized by the ratio $r_d'/r_s'$. 
}
\label{sphere_cap}
\end{figure}

We first discuss if the +1 defect can be supported by spherical geometry. All the
degenerate nematic configurations containing a +1 defect at the center of the
spherical cap are characterized by the director field $\mathbf{n}=(c_1,c_2)$,
where the constants $c_1$ and $c_2$ satisfy $c_1^2+c_2^2=1$. These degenerate states have
the same Frank free energy:
\begin{eqnarray}\label{}
  F_{+1,s} &=& \frac{K_1}{2} \iint_{\theta, \varphi \in D} \frac{1}{R^2
  \textrm{sin}^2 \theta} (R^2 \textrm{sin} \theta d\theta d\varphi) \nonumber \\
  &=&\frac{K_1}{2} \iint_{\theta, \varphi \in D} \frac{1}{\textrm{sin} \theta}
  d\theta d\varphi,\label{Fs_theta}
\end{eqnarray}
where the integration is over a
spherical cap $D$ with spherical radius $R$ and geodesic radius $r_s$. And
these states are solutions to the Euler-Lagrange equation (see Appendix C). 

In order to derive for
$F_{+1,s} - F_{+1,p}$, the free energy difference of a +1 defect configuration
on spherical and planar disks, we introduce the following coordinates transformation. For generality, the Cartesian coordinates
of the center of the spherical cap are $(c,0,\sqrt{R^2-c^2})$ as shown in Fig.~\ref{sphere_cap}. The
center of the spherical cap is located at the north pole for $c=0$. The region of the
spherical cap is $D = \{(x,y,z)|x^2+y^2+z^2=R^2,
(x-c)^2+y^2+(z-\sqrt{R^2-c^2})^2 \leq r_s'^2\}$. $r_s'$ is the Euclidean distance from the
center to the boundary of the spherical cap. The
area of such a spherical cap is $S = \pi r_s'^2$. Now we construct the stereographic
projection from the spherical cap to the plane of equator. Specifically, we draw
a line connecting the south pole of the sphere and any point at $(x,y,z)$ or $(\theta, \varphi)$ on the spherical cap.
The point on the spherical cap is thus projected
to the intersection point $(u, v)$ of this line and the equator plane.
The
projection is described by the formula
\begin{equation}\label{}
  (u, v) = (\frac{Rx}{z+R},\frac{Ry}{z+R}),
\end{equation}
or, in terms of spherical coordinates,
\begin{equation}
  (u,v) = (\frac{R \textrm{sin} \theta \textrm{cos} \varphi}{\textrm{cos} \theta
  + 1},\frac{R \textrm{sin} \theta \textrm{sin} \varphi}{\textrm{cos} \theta +1
  }).\label{uv_tp}
\end{equation}

The stereographic projection has a convenient geometric property that any
spherical cap not containing the point of projection (south pole) is projected to a circular disk on the
equator plane:
\begin{equation}\label{}
  (u-u_0)^2+v^2 \leq r_{eq}^2,\label{req}
\end{equation}
where
\begin{equation} u_0=\frac{2cR^2}{-r_s'^2+2R(R+\sqrt{R^2-c^2})}, \nonumber
\end{equation}
and
\begin{equation} r_{eq}^2
=\frac{r_s'^2R^2(4R^2-r_s'^2)}{[r_s'^2-2R(R+\sqrt{R^2-c^2})]^2}. \nonumber
\end{equation}
To guarantee that the spherical cap contains the north pole, it is required that
$r_s'^2 \geq 2R(R-\sqrt{R^2-c^2})$. Alternatively, $c \leq r_s'
\sqrt{1-[r_s'/(2R)]^2}$ for given $r_s'$.  On the other hand, the spherical cap occupies
no more than half of a sphere, so $r_s' \leq \sqrt{2}R$.

From the Jacobian of the coordinates transformation in Eq.(\ref{uv_tp})
\begin{align*}
\frac{\partial(u,v)}{\partial (\theta,\varphi)} =
\left(
\begin{array}{cc}
    \frac{R \textrm{cos} \varphi}{1 + \textrm{cos} \theta} & -\frac{R \textrm{sin} \theta \textrm{sin} \varphi}{1 + \textrm{cos} \theta} \\
    \frac{R \textrm{sin} \varphi}{1 + \textrm{cos} \theta} & \frac{R \textrm{sin} \theta \textrm{cos} \varphi}{1 + \textrm{cos} \theta} \\
\end{array}
\right),
\end{align*}
and
\begin{equation*}
  dudv =  \left| \frac{\partial(u,v)}{\partial (\theta,\varphi)} \right| d\theta d\varphi
  = \frac{u^2+v^2}{\textrm{sin} \theta} d\theta d\varphi,
\end{equation*}
we finally have
\begin{equation}
  \frac{dudv}{u^2+v^2}=\frac{d\theta d\varphi}{\textrm{sin} \theta}.
\end{equation}
We therefore obtain the desired expression for Eq.(\ref{Fs_theta}) in the $(u, v)$ coordinates:
\begin{equation}\label{}
  F_{+1,s} = \frac{K_1}{2} \iint_{D} \frac{1}{u^2+v^2} dudv,\label{Fs_uv}
\end{equation}
where the integral domain $D=\{(u, v)| u^2+v^2 \leq r_s'^2
R^2/(4R^2-r_s'^2) \}$. Note that now the integrands in Eq.(\ref{Fs_uv}) and
Eq.(\ref{Fp}) have the same functional form and can be conveniently
compared. A subtle point worth mentioning is that the direct subtraction of
Eq.(\ref{Fp}) from Eq.(\ref{Fs_uv}) will lead to a wrong expression of $\Delta F
= F_{+1,s} - F_{+1,p}   = - (\pi/2) K_1 \ln [4-(r_s'/R)^2]$. One
can check that $\Delta F$ fails to converge to the expected zero in the limit of
$R\rightarrow \infty$. Here, the subtlety is
from the fact that the integrands in Eq.(\ref{Fs_uv}) and
Eq.(\ref{Fp}) have singularity at the origin point. To eliminate this
singularity, one has to cut off the small defect core. The integral domain of
Eq.(\ref{Fs_uv}) should be $D=\{(u, v)| (a/2)^2 \leq u^2+v^2 \leq r_s'^2
R^2/(4R^2-r_s'^2) \}$, where $a$ is the radius of the defect core. The prefactor
of $1/2$ is due to the shrink of the defect size in the previously introduced stereographic projection. The integral
domain in Eq.(\ref{Fp}) also becomes $a^2 \leq x^2+y^2 \leq r_p^2$.
To conclude, the change of the total free energy in the
deformation of the planar to the spherical nematic disk in the constraint of
fixed disk area $A_d$ is
\begin{eqnarray}
  \Delta F &=& F_{+1,s} - F_{+1,p} \nonumber \\
           &=& - \frac{\pi}{2} K_1 \ln \left( 1-(\frac{A_d}{4\pi R^2}) \right).\label{DF}
\end{eqnarray}
We check that $\Delta F$ approaches zero in the limit of $R\rightarrow \infty$,
as expected. Equation (\ref{DF}) shows that $F_{+1,s}$ is always larger than
$F_{+1,p}$.

However, it will be shown that a +1 defect can be stabilized within a
sufficiently curved spherical disk despite the higher energy in comparison with
the planar disk case. We analyze the stability of the +1 defect from the
derivative of the free energy with respect to its position in the disk.  The
expression for the free energy is rewritten in the new coordinates $\{x, y \}$,
where $x=u-u_0$ and $y=v$:
\begin{eqnarray}
  F_{+1,s}(c) = \frac{K_1}{2} \iint_{x^2+y^2 \leq r_{eq}^2}
\frac{1}{(x+u_0)^2+y^2} dxdy \nonumber \\
       = K_1 \int _{-r_{eq}}^{r_{eq}} dx \frac{1}{x+u_0} \textrm{arctan}
       \frac{\sqrt{r_{eq}^2-x^2}}{x+u_0}, \label{Fsc}
\end{eqnarray}
where $u_0$ and $r_{eq}$ are given in Eq.(\ref{req}). From Eq.(\ref{Fsc}), we have
\begin{eqnarray}
F'_{+1,s}(c) = K_1 \int_{-r_{eq}}^{r_{eq}} ( G_1 + G_2 + G_3 )dx,
\end{eqnarray}
where $G_1=-[u_0'(c)/(x+u_0)^2] \textrm{arctan} [\sqrt{r_{eq}^2-x^2}/(x+u_0)]$,
$G_2=- u_0'(c) \sqrt{r_{eq}^2-x^2}/[(x+u_0)(u_0^2-2u_0 x +r_{eq}^2)]$, and
$G_3=r_{eq}r_{eq}'(c)/[(u_0^2-2u_0 x+r_{eq}^2)\sqrt{r_{eq}^2-x^2}]$.
The $G_3$ term can be integrated
out: $K_1 \int_{-r_{eq}}^{r_{eq}} G_3 dx= K_1 \pi r_{eq}
r_{eq}'(c)/|u_0^2-r_{eq}^2|$.
Local analysis around the defect at $x=-u_0$ shows that both the
$G_1$ and the $G_2$ terms are odd functions of $x$, and can be canceled in the
integration of $x$ near the defect. The singularity associated with the defect
is therefore removed. Note that $F'_{+1,s}(c)$ is negative in the large $R$ limit, which is consistent with the planar
disk case.

Now we analyze zero points of $F'_{+1,s}(c)$. The defect is stable at a zero
point where the slope of the $F'_{+1,s}(c)$ curve is positive.  With the
increase of $c$, numerical analysis shows that the $G_1$ term decreases and the
$G_3$ term increases, both starting from zero at $c=0$. While the $G_1$ and the
$G_3$ terms are comparable, the $G_3$ term is much smaller than either of them.
The competition of the $G_1$ and the $G_3$ terms may lead to
another zero point at the $F'_{+1,s}(c)$ curve in addition to the unstable zero
point at $c=0$.

\begin{figure}
\includegraphics[width=8.5cm]{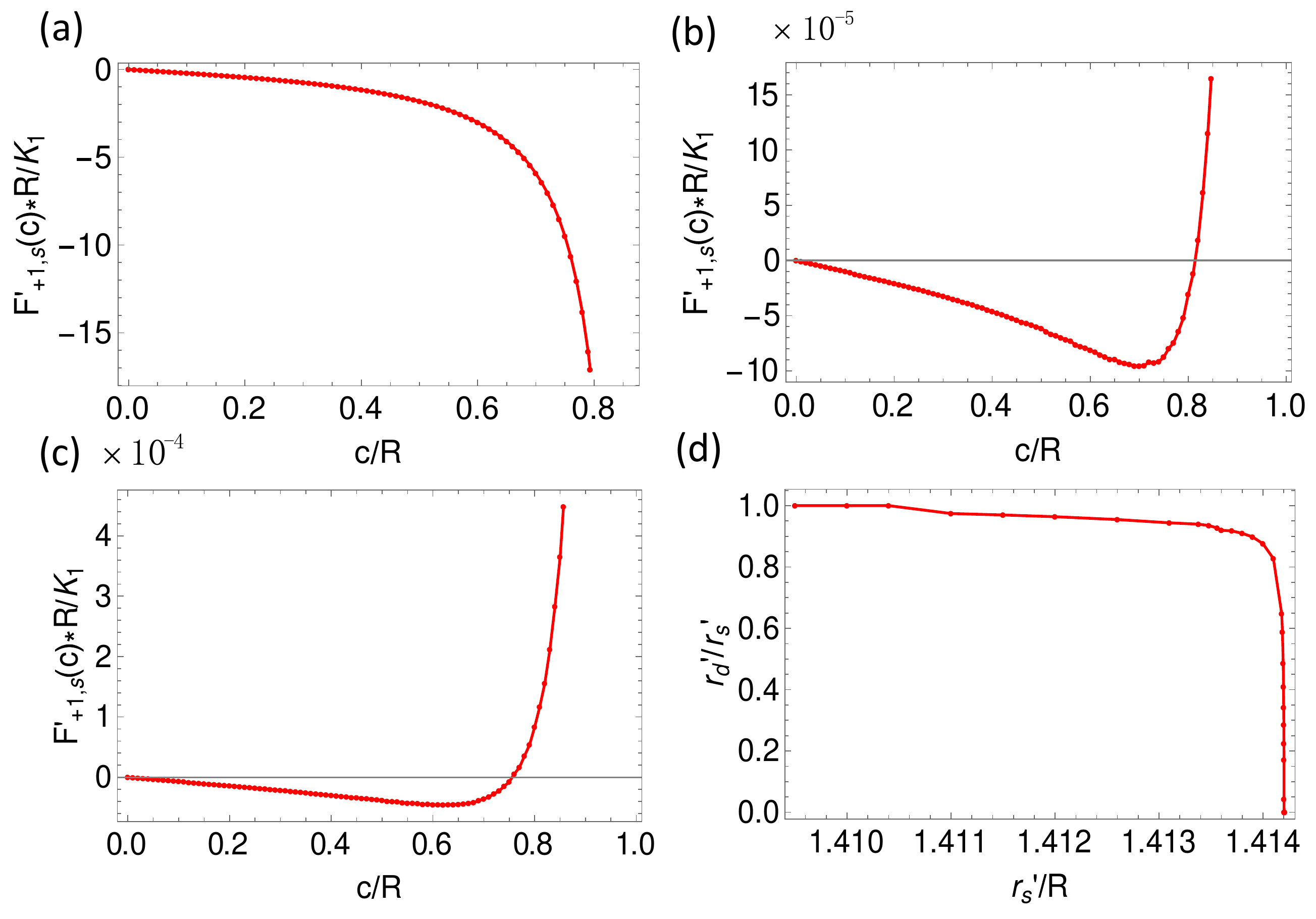}
\caption{Stability analysis of a +1 defect in a spherical nematic disk. (a)-(c)
show the derivative of the Frank free energy $F'_{+1,s}(c)$ versus $c$ at
typical values for $r_s'/R$. The curve starts to develop a stable zero point
for the +1 defect with the increase of the disk size $r_s'$. $r_s'/R=1.0$ (a),
1.414180 (b), and 1.414187 (c). (d) shows the rapid movement of the equilibrium
position of the +1 defect from the boundary ($r_d'/r_s'=1$) to the center
($r_d'/r_s'=0$) of the disk with the increase of the disk size $r_s'/R$. 
}
  \label{rdrs}
\end{figure}

In Figs.~\ref{rdrs}(a)--\ref{rdrs}(c), we plot $F'_{+1,s}(c)$ versus $c$ at typical values
for $r_s'$. We see that the $F'_{+1,s}(c)$ is negative and monotonously decreasing
when the spherical cap is smaller than a critical value.  With the increase
of $r_s'$, a second zero point appears at $c=c^*$, where a perturbed
defect will be restored to the original equilibrium position. It indicates that
the equilibrium position of the defect starts to depart from the boundary of the
disk. We introduce the quantity $r_d'/r_s'$ to characterize the equilibrium
position of the defect over the spherical cap, where $r_d'$ is the Euclidean
distance between the center of the disk and the defect. The variation of the
optimal position of the +1 defect with the size of the spherical cap is
summarized in Fig.~\ref{rdrs}(d). A pronouncing feature of the $r_d'/r_s'$ vs
$r_s'/R$ curve is the rapid decrease from unity to zero when $r_s'/R$ varies by
only about $0.1 \%$. It corresponds to the movement of the defect from the boundary
to the center of the disk. Such a transition occurs in the narrow window of
$r_s'$ when the spherical cap occupies about half of the sphere. Note that the
spherical cap becomes a half sphere when $r_s'=\sqrt{2} R$.

Here, it is of interest to compare a +1 defect in nematics and a five-fold
disclination in a two-dimensional hexagonal crystal on a sphere. Both nematic and
crystalline order are frustrated on a sphere, leading to the proliferation of
defects. The resulting defects in condensed matter orders are to screen the
geometric charge of the substrate surface, which is defined to be the integral
of Gaussian curvature. Over a spherical crystal, the topological charge of a
five-fold disclination can be screened by a spherical cap of area $A_0/12$
($A_0$ is the area of sphere), since 12 five-fold disclinations are required
over a spherical crystal by topological constraint~\cite{nelson2002defects}.
Topological analysis of a spherical nematics shows that a sphere can support
two +1 defects, so the topological charge of a +1 defect can be screened by a
spherical cap of area $A_0/2$. Our energetics calculation is consistent with
such topological analysis; it is when the spherical cap becomes as large as
a half sphere that a +1 defect will be energetically driven to move to the
center of the disk.

\begin{figure}
\includegraphics[width=8.5cm]{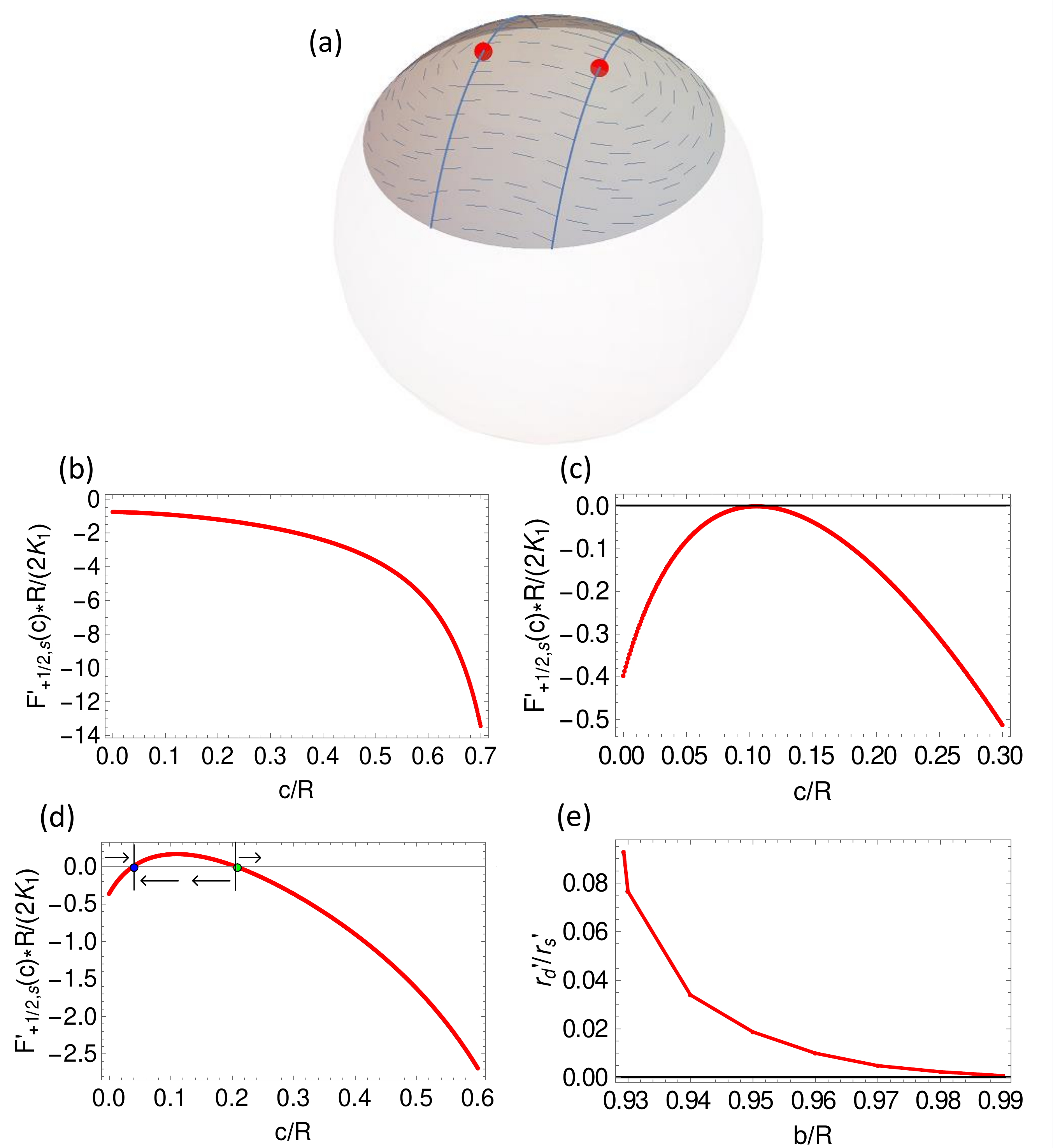}
\caption{Stability analysis of a pair of +1/2 defects in a spherical nematic
disk.  (a) is the schematic plot of the defect pair over the spherical disk in
gray. (b)-(d) show the derivative of the Frank free energy $F'_{+1/2,s}(c)$
versus $c$ at typical values for $b/R$. $b$ is the radius of the circular
boundary of the spherical cap. A pair of zero points appear in the curve with
the increase of the disk size $b$. $b/R=0.80000$ (b), 0.92933 (c), and 0.94000
(d). The curvature-driven alternating repulsive and attractive regimes in the
$F'_{+1/2,s}(c)$ curve are indicated by the arrows in (d). In (e), we plot the
variation of the equilibrium location of the defect pair versus the disk size
$b/R$. $r_d'$ is the Euclidean distance from one of the two defects to the
center of the disk. The defect pair merge to form a +1 defect ($r_d'/r_s'
\rightarrow 0$) in the half sphere limit ($b/R \rightarrow 1$).
}
  \label{defect_121}
\end{figure}

We proceed to discuss the split of a +1 defect into two +1/2 defects over a
spherical cap. Like the case of the planar disk, we first construct the director
field containing two +1/2 defects by cutting an azimuthal +1 defect
configuration. As shown in Fig.~\ref{defect_121}(a), the resulting director field on
the spherical cap is composed of three parts: the middle uniform region where
$\mathbf{n}=(-z/\sqrt{x^2+z^2},0,x/\sqrt{x^2+z^2})$, and the symmetric
azimuthal configurations at the two sides. The origin of the Cartesian coordinates is at the center of the sphere, and the z-axis passes through the north pole.  The two +1/2 defects are indicated by red dots in
Fig.5. Their x-coordinates are $x=\pm c$. The center of the spherical
cap is at the north pole.
The Frank free energy density of the middle uniform configuration is
$f=K_1/[2(x^2+z^2)]$.
By putting them together and working in the Cartesian coordinates over the equator plane, we have
\begin{eqnarray}
  \frac{F_{+ \frac{1}{2},s}}{2K_1}= \iint_{D_1} \frac{1}{x^2+y^2}dA+\iint_{D_2}\frac{1}{R^2-y^2}dA,   \label{Fc2}
\end{eqnarray}
where the surface element
of the spherical cap $dA=(R/\sqrt{R^2-x^2-y^2})dxdy$, $D_1 = \{(x, y)| y\in[0,
\sqrt{b^2-(x+c)^2}],  x\in[0, b-c] \}$, and $D_2 = \{(x, y)| y\in[0,
\sqrt{b^2-x^2}],  x\in[0, c] \}$. $b$ is the radius of the
circular boundary of the spherical cap. $b=r_s' \sqrt{1-(r_s'/2R)^2}$.

From Eq.(\ref{Fc2}), we have
\begin{eqnarray}
  \frac{F'_{+\frac{1}{2},s}(c)}{2K_1}=
\int_0^{\frac{b-c}{R}}F_1(x)dx+
  \int_0^{\sqrt{b^2-c^2}}F_2(y)dy, \label{Fc2d}
\end{eqnarray}
where $F_1(x)=R^2 (c+Rx)/\{\sqrt{-b^2+c^2+R^2+2Rcx} \\
  \sqrt{b^2-(c+xR)^2} [-b^2+c(c+2Rx)]\}$, and $F_2(y)=
  R/[(R^2-y^2)\sqrt{R^2-c^2-y^2}]$.
It is straightforward to show that $F'_{+1/2,s}(0)=-\sqrt{R^2-b^2}/(Rb) < 0$. It
indicates the repulsive interaction between two infinitely close +1/2 defects.
Numerical evaluation of
Eq.(\ref{Fc2d}) shows that when the spherical disk is sufficiently large, the
departing +1/2 defects can be stabilized within the disk. The plots of
$F'_{+1/2,s}(c)$ at typical values for $b/R$ are shown in
Figs.~\ref{defect_121}(b)--\ref{defect_121}(d). We see that when $b/R>0.93$, the
$F'_{+1/2,s}(c)$ curve starts to hit the horizontal zero line, leading to the
two zero points indicated by the blue and the green dots in
Fig.\ref{defect_121}(d). When the separation between the two defects is smaller
than the value at the blue dot or larger than the value at the green dot, they
repel with each other. In the regime between the two zero points, the defects
attract with each other. The curvature-driven alternating repulsive and
attractive regimes in the $F'_{+1/2,s}(c)$ curve are indicated by the arrows in
Fig.\ref{defect_121}(d). The left zero point (blue dot) 
 represents the equilibrium configuration of the +1/2
defects. In Fig.~\ref{defect_121}(e), we show the variation of the equilibrium
position of the +1/2 defects with the size of the spherical disk. When the
spherical cap occupies more area over the sphere, the distance between the two
+1/2 defects in the equilibrium configuration shrinks. In the limit of a
half sphere, the two +1/2 defects merge together, becoming a +1 defect.
This result is consistent with our previous analysis of the +1 defect case, where the
optimal position of the +1 defect over a half sphere is at the center of
the disk.

We proceed to discuss nematic order on $Poincar\acute{e}$ disk with
constant negative Gaussian curvature~\cite{hyperbolicKamien}. The associated metric over a hyperbolic
disk with Gaussian curvature $K_G$ is characterized by $ds^2 = 4(dx^2+dy^2)/(1+K_G
    r^2)^2$, where $r^2=x^2+y^2$. The area element
$dA=4dxdy/(1+ K_G r^2)^2$. For the director field $\textbf{n} = n_1(x,y)
  \mathbf{e}_1 + n_2(x,y)\mathbf{e}_2$, where $\mathbf{e}_1$ and $\mathbf{e}_2$
  are the orthogonal unit basis vectors, its divergence and curl are div~$\textbf{n} = (1/2)(1+ K_G r^2) (\partial n_1/\partial x +
      \partial n_2/\partial y) - K_G (n_1x + n_2y)$, and
  $\textrm{curl}\ \textbf{n} = (1/2)(1+ K_G r^2) (\partial
      n_2/\partial x - \partial n_1/\partial y) - K_G (n_2x -
	n_1y)$, respectively (see Appendix B for the derivation of curl $\textbf{n}$). 
We first consider a defect-free uniform director field $(n_1,n_2)=(\textrm{cos}
    \theta_0, \textrm{sin} \theta_0)$ whose associated Lagrange multiplier is
$\lambda=K_1 K_G (1+K_G r^2)$. $\theta_0 \in [0, \pi/2]$.  The associated Frank free
energy density is independent of $\theta_0$: $f = K_1 K_G^2r^2/2$. We see that the uniform state in $Poincar\acute{e}$ disk has a
non-zero energy density that increases with $r$ in a power law. It is due to the
special metric structure of the $Poincar\acute{e}$ disk.  Now we consider a +1 defect
configuration in the nematic texture on $Poincar\acute{e}$ disk. It can be
characterized by $\textbf{n}=[(c_2 x - c_1 y)\mathbf{e}_1+
(c_1 x + c_2 y) \mathbf{e}_2] /\sqrt{x^2+y^2}$,
where $c_1$ and $c_2$ are both constants satisfying $c_1^2+c_2^2=1$, such that
the magnitude of $\mathbf{n}$ is unity. Varying the value of $c_1$ from zero to
unity, we obtain director fields from radial to azimuthal configurations.
The associated Frank free energy density is
  \begin{align}\label{}
  f_{+1,h} = \frac{K_1}{2} \frac{(1- K_G r^2)^2}{4r^2}.
  \end{align}
Since $f'_{+1,h}(r) = (K_1/2) (-1+K^2 r^4)/(2r^3) < 0$, the Frank free energy
density decreases with $r$. On the other hand, due to the homogeneity of the $Poincar\acute{e}$
disk, the optimal position of a +1 defect is always at the
boundary of the disk.

Finally, we discuss some effects that are not taken into consideration in our
calculations.  First, by introducing anisotropy in the elastic constants, the
free energy varies with the local rotation of the director field. Despite the
reduced energy degeneracy arising from the elasticity anisotropy, both radial and azimuthal configurations based on
which our calculations are performed are still solutions to the Euler-Lagrange
equation (see Appendix C). Therefore, introducing elasticity anisotropy does
not change the major conclusions about the optimal positions of
both $+1$ and $+1/2$ defects. Second, in addition to curvature, the thickness of
liquid-crystal
shells is an important parameter to control the number and orientation of
defects~\cite{lopez2011frustrated,koning2016spherical}. It has been
experimentally observed that thickness variation can produce a number of novel
defect configurations over a spherical liquid-crystal
shell~\cite{lopez2011frustrated}.  It is of great interest to include the
effect of thickness in a generalized Frank free energy model to account for
these new experimental observations~\cite{koning2016spherical}.  This is beyond
the scope of this study.  Third, spatial variations in nematic order parameter
within defect cores contribute to the condensation free energy of topological
defects~\cite{kleman1983points, chaikin2000principles}. Notably, nematic
textures in defect core regions can exhibit featured patterns and energy
profiles, such as highly biaxial nematic order in the cores of $+1/2$
defects~\cite{schopohl1987defect} and local melting of the nematic
ordering~\cite{kralj2011curvature}. A recent study has demonstrated that the
condensation energy associated with the defect core plays an important role in
the formation of defects triggered by strong enough
curvature~\cite{mesarec2016effective}. In our study, we focus on the optimal
locations of pre-existent defects. They are determined by the variation of the
free energy with the positions of the defects, where the contribution from the
defect core structures is canceled without considering the boundary effect of
defects.

\section{CONCLUSION}

In summary, we investigate the curvature-driven stability mechanism of LC
defects based on the isotropic nematic disk model where the appearance of
defects is not topologically required, and present analytical
results on the distinct energy landscape of LC defects created by curvature. We
show that with the accumulation of curvature effect both +1 and +1/2 defects
can be stabilized within spherical disks. Specifically, the equilibrium
position of the +1 defect will move abruptly from the boundary to the center of
the spherical disk, exhibiting the first-order phase-transition-like
behavior.  We also find the alternating repulsive and attractive regimes in the
energy curve of a pair of +1/2 defects, which leads to an equilibrium defect
pair separation. The sensitive response of defects to curvature and the
curvature-driven stability mechanism demonstrated in this work may have
implications in the control of LC textures with the dimension of curvature.

\section*{Appendix A: Infinite degree of degeneracy in the one elastic constant
approximation}

Let us consider a planar nematic disk.  $\mathbf{n} = (\textrm{cos}
    \theta(x,y),\textrm{sin} \theta(x,y))$ in Cartesian coordinates. In the one elastic constant
approximation, the Frank free energy is $F=\frac{1}{2}K_1 \iint_ {x^2+y^2 \leq
  r_p^2} |\triangledown \theta(x,y)|^2 dxdy$. It is easily seen that rotating a
  director by a constant angle does not change the Frank free energy.

For a nematic field on a sphere, by inserting  $n=(\textrm{cos}
    \Psi(\theta,\varphi),\textrm{sin} \Psi(\theta,\varphi))$ in spherical coordinates into
Eq.(\ref{Fsphere}), we obtain the expression for the Frank free energy density
\begin{align}\label{}
  f=\frac{K_1}{2}(\frac{1}{R^2 \textrm{sin}^2 \theta}+ |\nabla \Psi|^2
      +\frac{2\textrm{cos} \theta}{R^2 \textrm{sin}^2 \theta}
      \Psi_{\varphi}),\nonumber
\end{align}
where $\nabla \Psi=\frac{1}{R}\frac{\partial \Psi}{\partial
  \theta}\textbf{e}_{\theta}+\frac{1}{R\textrm{sin} \theta}\frac{\partial
    \Psi}{\partial \varphi}\textbf{e}_{\varphi}$, and the notation
    $\Psi_{\varphi}$ is an abbreviation for $\partial \Psi / \partial \varphi$.
Obviously, the Frank free energy density is invariant under the transformation $\Psi \rightarrow \Psi+c$.

The conclusion that the nematic texture has infinite degree of
degeneracy in the one elastic constant approximation can be generalized to any
generally curved surface by writing the Frank free energy under the one
constant approximation in the form of
\begin{eqnarray}
F=\frac{1}{2}\int dS g^{ij}(\partial_i \alpha - A_i) (\partial_j \alpha - A_j),
  \nonumber
\end{eqnarray}
where the integration is over an area element $dS$ on the surface $\mathbf{x}(u^1,
    u^2)$, $\alpha(u^1,u^2)$ is the angle between $\mathbf{n}(u^1,u^2)$ and any local
reference frame, and $A_i$ is the spin connection~\cite{vitelli2004defect}. The free
energy is invariant under the rotation $\alpha(u^1,u^2)\rightarrow
\alpha(u^1,u^2) +c$.

\section*{Appendix B: Calculating curl $\mathbf{n}$ on spherical geometry}

In a coordinates independent expression, curl~$\mathbf{n}$=$(\star d n^{\flat})^{\sharp}$~\cite{Nakahara,panoramic,multivariable}.
The operators $\star$, $\scriptstyle\flat$ and $\scriptstyle\sharp$ are to be
explained below. $\star$ is an operator called Hodge dual. When applied on an antisymmetric tensor
$\alpha=\frac{1}{k !} \alpha_{i_1,\cdot\cdot\cdot,i_k} e^{i_1} \wedge
\cdot\cdot\cdot \wedge e^{i_k}$, where $e^{i_1}, \cdot\cdot\cdot, e^{i_n}$
are dual bases,
    \begin{eqnarray}
    \star \alpha = \frac{\sqrt{|g|} \varepsilon_{i_1,
  \cdot\cdot\cdot, i_n}  \alpha_{j_1,\cdot\cdot\cdot,j_k}g^{i_1j_1}
  \cdot\cdot\cdot g^{i_kj_k} }{k!(n-k)!}  e^{i_{k+1}}\wedge \cdot\cdot\cdot \wedge
  e^{i_n}.\nonumber
  \end{eqnarray}
$\scriptstyle\flat$ and $\scriptstyle\sharp$ are the musical isomorphisms. $X^{\flat}=g_{ij}X^i dx^j$, where
$X=X^{i}\partial_i$. $\omega^{\sharp}=g^{ij}\omega_i \partial_j$, where
$\omega=\omega_i dx^i$.

Consider a vector field $\mathbf{n}$ defined on a two-dimensional sphere. $\mathbf{n} = n_1(\theta,\varphi)\mathbf{e}_{\theta}
+ n_2(\theta,\varphi)\mathbf{e}_{\varphi}$, where $\mathbf{e}_{\theta}$ and $\mathbf{e}_{\varphi}$ are the unit tangent
vectors in spherical coordinates. Applying the above formulas on such a vector
field, we have 
\begin{eqnarray}
n^{\flat } &=& n_{1}rd\theta +n_{2}r\sin \theta d\varphi, \nonumber
\end{eqnarray}
\begin{eqnarray}
dn^{\flat } &=& n_{1}drd\theta +n_{2}\sin \theta drd\varphi \nonumber \\ 
&+& r(n_{2}\cos \theta +\sin \theta \frac{\partial n_{2}}{\partial \theta }-
\frac{\partial n_{1}}{\partial \varphi })d\theta d\varphi, \nonumber
\end{eqnarray}
and
\begin{eqnarray}
\star dn^{\flat } &=& n_{1}\sin \theta d\varphi -n_{2}d\theta \nonumber \\
&+& \frac{1}{r\sin \theta }(n_{2}\cos \theta +\sin \theta \frac{\partial n_{2}
}{\partial \theta }-\frac{\partial n_{1}}{\partial \varphi })dr. \nonumber
\end{eqnarray}

We finally obtain Eq.(\ref{curl_sphere}).
It is of interest to note that the curl of a director field $\mathbf{n}$ on a
generally curved surface is $\textrm{curl}\ \mathbf{n} = -\tau_{n} \mathbf{n}
-c_{n} \mathbf{t} + \kappa_{n} \boldsymbol{\nu}$, where
$\{\mathbf{n}, \mathbf{t}, \boldsymbol{\nu}  \}$ constitute the Darboux
basis~\cite{napoli2012extrinsic}. $\tau_{n}$ and $c_{n}$ are
the components of the extrinsic curvature tensor $\mathbf{L}$.
$\mathbf{L}_{nn}=c_n$, and $\mathbf{L}_{nt}=\mathbf{L}_{tn}=-\tau_n$. In
general, the extrinsic curvature influences the Frank free energy of nematics
on a curved surface. It is only on a flat or spherical surface 
  $\tau_{n}=0$ and $c_{n}$ is a constant. So the
  extrinsic curvature effect only contributes a constant term in the Frank
  free energy~\cite{napoli2012extrinsic}.

\section*{Appendix C: Euler-Lagrange equations in Cartesian and spherical
    coordinates}

In this appendix, we present the Euler-Lagrange equations in Cartesian and spherical coordinates
derived from Eq.(\ref{EL_general}), and show that both radial and azimuthal
configurations are solutions to the Euler-Lagrange equations.
We also show that the anisotropic elastic constants will not change
the main result of curvature-driven alternating repulsive and attractive
interactions between the two +1/2 defects due to the fact that the elastic modulus $K_1$ plays no role in the energy expression.

In two-dimensional Cartesian coordinates,
$\mathbf{n}=[n_1(x,y),n_2(x,y)]$. The components of the director field in
equilibrium nematic textures satisfy the
following Euler-Lagrange equations:
\begin{eqnarray}
K_1 (\frac{\partial^2 n_1}{\partial x^2} + \frac{\partial^2 n_2}{\partial x
    \partial y}) - K_3(\frac{\partial^2 n_2}{\partial x \partial y} -
      \frac{\partial^2 n_1}{\partial y^2 })=-\lambda n_1,\nonumber
    \end{eqnarray}
and
\begin{eqnarray}
K_1 (\frac{\partial^2 n_2}{\partial y^2} + \frac{\partial^2 n_1}{\partial x
    \partial y}) + K_3(\frac{\partial^2 n_2}{\partial x^2} - \frac{\partial^2
      n_1}{\partial x \partial y })=-\lambda n_2.\nonumber
\end{eqnarray}
It is found that both radial $(n_1,n_2)= (x/\sqrt{x^2+y^2},y/\sqrt{x^2+y^2})$
 and azimuthal $(n_1,n_2) = (-y/\sqrt{x^2+y^2},x/\sqrt{x^2+y^2})$
    configurations satisfy the above Euler-Lagrange equations with $\lambda
    =K_1/(x^2+y^2)$ and $\lambda =K_3/(x^2+y^2)$, respectively.
The spiral configuration $\mathbf{n}=c_1(x/\sqrt{x^2+y^2},y/\sqrt{x^2+y^2})+c_2(-y/\sqrt{x^2+y^2},x/\sqrt{x^2+y^2})$ ($c_1^2+c_2^2=1$ and neither $c_1$ nor $c_2$ is 0) is the solution to the
Euler-Lagrange equations only in the one elastic constant approximation.

In spherical coordinates,
   $\mathbf{n}=n_1\mathbf{e}_{\theta}+n_2\mathbf{e}_{\varphi}$. In
equilibrium nematic textures, $n_1$ and $n_2$ satisfy the
following Euler-Lagrange equations: 
\begin{align*}\label{}
   & \frac{K_1}{R^2}(-\frac{n_1}{\textrm{sin}^2 \theta}+\frac{\partial n_1}{\partial \theta} \frac{\textrm{cos} \theta}{\textrm{sin} \theta}+\frac{\partial^2 n_1}{\partial \theta^2}-\frac{\textrm{cos} \theta}{\textrm{sin}^2 \theta} \frac{\partial n_2}{\partial \varphi} \\
   & +\frac{1}{\textrm{sin} \theta}\frac{\partial^2 n_2}{\partial \theta
   \partial \varphi})+\frac{K_3}{r^2 \textrm{sin}^2 \theta}(\frac{\partial^2 n_1}{\partial \varphi^2}-\frac{\partial n_2}{\partial \varphi} \textrm{cos}\theta \\
  & -\textrm{sin} \theta \frac{\partial^2 n_2}{\partial \theta \partial
  \varphi}) = -\lambda n_1, \\
  & \frac{K_3}{R^2}(-\frac{ n_2}{\textrm{sin}^2 \theta} + \frac{ \textrm{cos} \theta}{ \textrm{sin}^2 \theta} \frac{\partial n_1}{\partial \varphi} - \frac{1}{\textrm{sin} \theta} \frac{\partial ^2 n_1}{\partial \theta \partial \varphi} + \frac{\partial^2 n_2}{\partial \theta^2} \\
  & + \frac{\textrm{cos} \theta}{\textrm{sin} \theta}\frac{\partial
  n_2}{\partial \theta})+\frac{K_1}{R^2 \textrm{sin} \theta} (\frac{\textrm{cos} \theta}{\textrm{sin} \theta}\frac{\partial n_1}{\partial \varphi}+\frac{\partial^2 n_1}{\partial \theta \partial \varphi} \\
  & +\frac{1}{\textrm{sin} \theta}\frac{\partial^2 n_2}{\partial
  \varphi^2}) = -\lambda n_2.
 \end{align*}
We remark that the equilibrium
equations in spherical coordinates are invariant under uniform local rotation of the
director field. Similarly, one can show that both radial and azimuthal configurations are
solutions to the Euler-Lagrange equations with $\lambda =K_1/(R^2
    \textrm{sin}^2 \theta)$ and $\lambda =K_3/(R^2 \textrm{sin}^2 \theta)$,
respectively.  The spiral configuration of $n_1=c_1, n_2=c_2$ ($c_1$ and
    $c_2$ are non-zero constants satisfying $c_1^2+c_2^2=1$) satisfies the
equilibrium equation only when $K_1 = K_3$. For the two $+1/2$ defects
configurations discussed in the main text, we show that introducing elasticity
anisotropy does not change the curvature-driven alternating repulsive and
attractive interactions between the  defects. For the two
$+1/2$ defects configuration on a spherical disk where an azimuthal
configuration is separated by a uniform configuration, the associated Frank
free
energy is \begin{align*} \frac{F_{+\frac{1}{2},s}}{2K_3}=\iint
_{D_1}\frac{1}{x^2+y^2}dA+\iint _{D_2}\frac{1}{R^2-y^2}dA, \end{align*} where
$D_1$ and $D_2$ are given below Eq.(\ref{Fc2}). We see that since the entire defect
configuration is divergence free, the parameter $K_3$ does not appear in the
expression for the Frank free energy. Therefore, anisotropy in elastic constants
does not change the featured interaction between the $+1/2$ defects.

\section*{Acknowledgement}

This work was supported by NSFC Grant No. 16Z103010253, the SJTU startup fund
under Grant No. WF220441904, and the award of the Chinese Thousand Talents
Program for Distinguished Young Scholars under Grant No. 16Z127060004.

\end{document}